\documentclass[onecolumn]{aa}

\usepackage{graphicx}
\def\ls{{_<\atop^{\sim}}}
\def\gs{{_>\atop^{\sim}}}
\def\cgs{ ${\rm erg~cm}^{-2}~{\rm s}^{-1}$ } 

\begin{document}
%%%%%%%%%%%%%%

\title{The X-ray absorber of PKS2126-158}

\author{Fabrizio Fiore\inst{1}, Martin Elvis\inst{2}, 
Roberto Maiolino\inst{3}, Fabrizio Nicastro\inst{2}, 
Aneta Siemiginowska\inst{2}, Giulia Stratta\inst{1,4} 
and Valerio D'Elia\inst{1}}

\institute {INAF - Osservatorio Astronomico di Roma \\
via Frascati 33, Monteporzio-Catone (RM), I00040 Italy.\\
\email{fiore@mporzio.astro.it,stratta@mporzio.astro.it,delia@mporzio.astro.it}
\and 
Harvard-Smithsonian Center for Astrophysics, 
Cambridge MA 02138, USA. 
\email{elvis@head-cfa.harvard.edu, aneta@head-cfa.harvard.edu}
\and 
INAF - Osservatorio Astrofisico di Arcetri.
\email{maiolino@arcetri.astro.it}
\and 
Universit\'a di Roma ``La Sapienza''.}

\date{03 June 2003}

\abstract{

BeppoSAX observed the z=3.27 quasar PKS2126-158 on 1999 May 24-28 when
its 2-10keV and 0.1-2.5keV fluxes were $1.1\times10^{-11}$ and
$4.4\times10^{-12}$ \cgs respectively, a factor of 2 higher than in
all previous ROSAT and ASCA observations and 40\% higher than in two
more recent Chandra and XMM-Newton observations. The shortest detected
rest frame variability timescale is of a few months, comparable to the
causal timescale associated to an emission region of $\sim10$
Schwarzschild radii around a few$\times10^{10}$M$_{\odot}$ black hole.
The source is detected with a signal to noise ratio $S/N\gs3$ up to
$\sim$50 keV, 215 keV rest frame. The BeppoSAX observations confirm
the presence of low energy absorption along the line of sight,
independent on the continuum model adopted, at high confidence level.
Despite the limited spectral resolution of the BeppoSAX LECS and MECS
it is possible to put constraints on different absorption and
continuum models, but not to unambiguously determine the redshift of
the absorber. If the absorber is not significantly ionized the
BeppoSAX data do prefer an absorber at z$\ls2.7$.  Strong and complex
metal line systems along the line of sight to PKS2126-158 have been
found at z=0.6631 and at 2.64$<$z$<$2.82.  They could well be
associated to the X-ray absorption.  Conversely, an ionized (``warm'')
absorber at the quasar redshift provides a good fit only if the iron
abundance is smaller than $\sim0.3$ solar, while that of the other
elements is fixed to the solar value.  Lower iron metallicity would
imply a lower dust to gas ratio, since iron aggregates easily in
dust. This can help in solving the apparent paradox of the lack of
significant ultraviolet reddening in this source while strong
absorption is detected in X-rays.  Low iron abundance would be at odds
with the supersolar abundances derived from the broad emission lines.


}

\authorrunning {Fiore et al.}
\titlerunning {The X-ray absorber of PKS2126-158}

\maketitle

%%%%%%%
\section{Introduction}

The high redshift (z$>$3) Universe can be probed by observing distant
convenient ``lighthouses'' such as quasars, Gamma Ray Bursts, and
luminous galaxies. Both emission and absorption features superimposed
onto the source continuum can be used for this purpose.  High quality
spectra of high redshift cosmic sources can tell us about the physics
and chemistry of the matter along the line of sight (i.e. density,
physical state and element abundances).  In particular, high-z quasars
and galaxies give a unique chance of probing the metallicity of the
gas enriched by the first generations of stars. Iron is expected to be
produced mostly by type Ia SNe, occurring with a delay of $\approx1$
Gyr with respect to the first type II SNe, which produce most of the
$\alpha$ elements such as O, Si, Mg.  Therefore, the relative
abundances of various elements produced on different timescale in
high-z quasars (e.g. the [$\alpha$/Fe] logarithmic ratio, where the
columns are normalized to their solar value) is a sensitive clock of
star-formation history, and its determination provides unique
constraints on the star-formation and on the initial mass function at
high redshift (e.g. Hamann \& Ferland 1999).

So far metallicities at high-z have been studied by means of a few
quasar and radio-galaxy UV rest-frame emission lines (see
e.g. Hamann \& Ferland 1999, Constantin et al. 2002, Hamann et al. 2002),
and via optical-UV absorption systems, both associated with the quasar and
with galaxies which happen to lie along the line of sight to bright
quasars. 

Emission line metallicity determinations are plagued by uncertainties
on the properties of the emitting gas (temperature, density, ion
fraction, dust depletion) and by the small number of lines available
to constrain the metallicity. In particular, the abundance of iron
relative to the $\alpha$ elements is mostly inferred from the
intensity of the UV Fe bump relative to MgII~(2803\AA ) (Thompson et
al. 1999, Iwamuro et al. 2002, Dietrich et al. 2002).  However, the
FeII UV bump is one of the most important coolants of the transition
region of the BLR in AGNs and therefore might be characterized by a
self-regulating intensity; indeed, Verner et al. (1999) have shown
that the emissivity of the Fe UV bump is far from having a 1:1
relation with the iron abundance. As a consequence, the iron
metallicity of high-z QSOs inferred by a naive comparison of the
FeII-UV/MgII intensity at high and low redshift might be quite
inaccurate or even deceiving.

Quasar intrinsic broad absorption line (BAL) metallicity
determinations provide bizarre but usually supersolar metal abundances
(Hamann 1998, Hamann \& Ferland 1999, Arav et al. 2001, Hamann et
al. 2003).  However these could have be overestimated, because strong
absorption lines like CIV could be optically thick and strongly
saturated (e.g. Hamann et al. 2003).  The analysis of BALs is
complicated by the fact the usually the line troughts do not reach
zero intensity, because the absorbers are leacky and/or because of
scattered/emitted flux from the absorbing region. Furtheremore, from the
optical/UV BAL it is not possible to measure the abuncances of iron,
which is one of the key elements to constrain the star formation history.

In the case of absorption features not associated with the
quasars, optical-UV metal line systems in galaxies along the line of
sight to quasars probe mainly the ISM of outer haloes, rather than the
bulge or the disk (see e.g. Churchill et al. 2000).  Even the highest
column density systems known, the Damped Ly$\alpha$ systems (DLAs),
may give a distorted view of the bulk of the `typical galaxy'
population at z=2-4.  DLA studies are likely to be biased against
dusty systems, since these would hide the background quasars in
color-based surveys (e.g.  Savaglio et al. 2000). Furthermore, in DLAs
metallicities are usually evaluated using faint zinc lines, since
iron, which is 20 times more abundant than zinc, is heavily depleted
onto dust grains (see e.g. Pettini et al. 1997). Using zinc lines
Pettini et al (1999) and Prochaska \& Wolfe (1999) estimate a
metallicity of about 0.1 at z=3-4. DLA metallicities may approach
solar values at low-z (Savaglio et al. 2000).

Summarizing, metallicity measurements at high redshift based on
optical observations have often provided only partial or uncertain
information. This statement applies specifically to iron. In contrast,
X-ray measurements offer a unique means of investigating absorption
systems because the X-ray absorption edges can constrain tightly
the ionization state, the molecular state and the dust content of
intervening material and so probe total column densities.  A direct
measurement of the depth of iron X-ray absorption edges will remove
the degeneracy involving various physical parameters of the emitting
region, which plague current metallicity estimates.  This can be
achieved at z=3-4 with the present instrumentation, because Fe-K
features at these redshift are redshifted to 1.5-2 keV, where the
X-ray instrument sensitivity is the greatest.

This approach is plausible as there are strong signs of absorption
toward several z=2-3 radio-loud quasars with implied column densities
of $N_H\sim10^{22}$ cm$^{-2}$ in the quasar rest frame (Elvis et
al. 1994, Serlemitsos et al. 1994, Cappi et al. 1997, Fiore et
al. 1998, Reeves \& Turner 2000).  The quality of the ROSAT and ASCA
discovery spectra however can not discriminate between a location for
the absorber at the quasar or along the intervening line of
sight. Fiore et al. (1998) and Elvis et al. (1998) argue for intrinsic
origin because of the lack of absorbers in radio-quiet quasars at
z=2-3. Kuhn et al. (2001) show that z=3 radio-quiet quasars have
spectra that are as UV-bright as low redshift radio-quiet quasars,
implying $A_V\ls0.3$ mags.  However the absence of dust and of
detectable X-ray $N_H$ in z=2-3 radio quiet quasars could be a
selection effect.  Most quasars are selected via optical colors and a
fairly modest reddening can drop the objects out of the samples.
Furthermore, radio-quiet quasars are fainter in X-rays than radio-loud
quasars and so have more poorly constrained spectra. So the issue of
the location of the absorbers in radio-loud quasars is still open.

PKS2126-158 is the brightest quasar in X-rays at z$>3$, and it is the
second brightest, after PKS2149--306, at z$>2$.  For this reason
PKS2126-158 has been studied in great detail at X-ray frequencies
since the first {\it Einstein} detection (Worrall \& Wilkes
1990). Elvis et al. (1994) discovered a strong low energy cut-off,
which is naturally explained by photoelectric absorption along the
line of sight, in addition to the Galactic 21cm column density of
$4.85\times10^{20}$ cm$^{-2}$ (Elvis et al. 1989).  The implied 
column density ranges from 
$0.9-20\times10^{21}$ cm$^{-2}$, if the absorber is at
z$\sim0$, it is not highly ionized and has solar abundances, up to a
column of $0.8-2.7\times10^{22}$ cm$^{-2}$ if the absorber is at
z$\sim3.27$, the quasar redshift.  ASCA observations confirmed the low
energy cutoff (Serlemitsos et al 1994, Cappi et al. 1997, Reeves \&
Turner 2000) but cannot distinguish beween genuine photoelectric
absorption or continuum curvature.  We have taken advantage of the
good BeppoSAX sensitivity over a very broad band, from 0.1 keV up to
200 keV, to disentangle photoelectric absorption and continuum
curvature and to constrain the redshift,  ionization state and
metallicity of the absorber in PKS2126-158.

The paper is organized as follows: Section 2 presents the observations
and data reduction; Section 3 presents the spectral analysis; Section
4 discusses the main results on the low energy absorption.

\section{BeppoSAX data reduction and analysis}

BeppoSAX observed PKS2126-158 from 1999 May 24 through 1999 May 28.
The observations were performed with the BeppoSAX Narrow Field
Instruments, LECS (0.1-10 keV, Parmar et al. 1997), MECS (1.3-10 keV,
Boella et al. 1997b), HPGSPC (4-60 keV, Manzo et al. 1997) and PDS
(13-200 keV, Frontera et al. 1997).  LECS and MECS are imaging gas
scintillation proportional counters, the HPGSPC is a collimated high
pressure gas scintillation proportional counter and the PDS consists
of four phoswich units. The PDS is operated in the so called ``rocking
mode'', with a pair of units pointing to the source while the other
pair monitor the background $\pm210$ arcmin away. The units on and off
source are interchanged every 96 seconds. We report here the analysis
of the LECS, MECS and PDS data; the HPGSPC data are very noisy, due to
the high HPGSPC background.  The MECS observations were performed with
units 2 and 3 (on 1997 May 6$^{\it th}$ a technical failure caused 
unit MECS1 to be switched off); these data were combined together after
gain equalization. The LECS was operated during dark time only,
therefore LECS exposure times are usually smaller than MECS ones.
Table 1 gives the LECS, MECS and PDS exposure times and the count
rates.

Standard data reduction was performed using the SAXDAS software
package version 2.0 following Fiore, Guainazzi and Grandi (1999).  In
particular, data are linearized and cleaned of Earth occultation
periods (we accumulated data for Earth elevation angles $>5$ degrees)
and unwanted periods of high particle background (satellite passages
through the South Atlantic Anomaly and periods with magnetic cut-off
rigidity $>6$ GeV/c). The low and stable particle background of the
LECS, MECS and PDS (with variations at most of 30\% around the orbit, due
to the low inclination of the satellite orbit, 3.95 degrees), ensures
that systematic errors in the subtraction of the background are
negligible, i.e. smaller, in the energy bands of interest, than the
statistical errors.

LECS and MECS spectra were extracted from regions of 8 arcmin and 3
arcmin radii respectively. These radii maximize the signal-to-noise
ratio below 1 keV in the LECS and above 2 keV in the MECS. Background
spectra were extracted in detector coordinates from high Galactic
latitude `blank' fields (98\_11 issue) using regions equal in size to
the source extraction region. We have compared the mean level of the
background in the LECS and MECS ``blank fields'' observations to the
mean level of the background in the PKS2126-158 observations using
source free regions at various positions in the detectors. The
``local'' MECS background count rate is within 2\% that in the ``blank
fields''. On the other hand, for the LECS the ``local'' background
count rate is 10\% smaller than that of the ``blank fields''.  For
this reason, the ``blank fields'' background spectrum was multiplied
by a factor of 0.9 before subtraction from the source spectrum.

The PDS data were reduced using the ``variable risetime
threshold'' technique to reject particle background (see Fiore,
Guainazzi \& Grandi 1999). This technique reduces the total 13-200 keV
background to about 20 counts s$^{-1}$ and the 13-80 keV background to
about 6 counts s$^{-1}$ (instead of 30 and 10 counts s$^{-1}$
respectively, obtained using the standard ``fixed risetime threshold''
technique).

The PDS rocking mode provides a reliable background subtraction.  This
can be checked by looking at the spectrum between 200 and 300 keV,
where the effective area of the PDS to X-ray photons is small and
therefore the source contribution is negligible. After background
subtraction we obtain a count rate of -0.0014$\pm$0.0100 counts
s$^{-1}$, fully consistent with the expected value of 0.  The net
(background subtracted) 13-100 keV on-source signal is 0.157$\pm$0.021
counts s$^{-1}$.  The source is detected with a signal to noise ratio
$S/N\gs3$ up to $\sim$50 keV, or 215 keV in the rest frame).
Confusion in the PDS collimator Field Of View (1.4 degrees FWHM)
ultimately limits our capability to constrain the high energy
spectrum. We have carefully checked for any possible contaminant in a
region of 1.5 degrees radius around the source (using the NED, SIMBAD,
AGN, clusters, CVs, Radio and X-ray sources catalogs) finding no
obvious bright hard X-ray source.  Of course there is the possibility
of a bright ``unknown'' source in the PDS field of view.  The chance
of finding a source in any given 2~square degrees, the PDS beam area,
is however small.  The HEAO-1 A4 all sky catalog (Levine et al. 1984)
lists just 7 high Galactic latitude sources in the 13-80~keV band down
to a flux of 2$\times$10$^{-10}$ \cgs (10mCrab). The 13-80 keV flux is
about 20 times smaller than this figure and so, assuming a logN-logS
slope of $-$1.5, we expect a chance coincidence rate of 2\%.

\begin{table}[ht]
\caption{\bf Observation log}
\begin{tabular}{lcc}
\hline
\hline
Instrument & Exposure (ks) & Count rate $^a$\\
\hline
LECS & 78.1  & 0.1-4 keV  5.50$\pm$0.09  \\
MECS & 107.9 & 1.7-10 keV 11.9$\pm$0.10  \\
PDS  & 51.6  & 13-100 keV 15.7$\pm$0.21 \\
\hline
\end{tabular}

$^a$ $10^{-2}$ counts s$^{-1}$
\end{table}

Spectral fits were performed using the XSPEC 11.0.1 software package
and public response matrices as from the 1999 December release.  For
the MECS and PDS we used the standard on axis matrices (effective area
and response matrix).  For the LECS we built the effective
area file corresponding to the source position in raw coordinates,
since it turned out that this position is relatively
far from the default pointing position (by about 3 raw pixels, i.e. 45
arcsec.) This caused the source to be closer than usual to one of the
tungsten wires supporting the LECS window.  Our LECS effective area
file takes than correctly into account the photons absorbed by the wire.

LECS and MECS spectra were rebinned following two criteria: a) to
sample the energy resolution of the detectors with four channels at
all energies whenever possible, and b) to obtain at least 20 counts
per energy channel.

Constant factors have been introduced in the fitting models in order
to take into account the intercalibration systematics between the
instruments (Fiore, Guainazzi and Grandi 1999). Assuming the MECS as
reference instrument, the expected factor between LECS and MECS is
about 0.9 [0.7-1.1].  Fitting the LECS and MECS spectra in the common
1.7-4 keV band with a simple model produced a factor of
0.835$\pm0.025$. In all following fits the LECS-MECS factor is
therefore constrained to vary between 0.81 and 0.86.

The expected factor between the PDS and MECS is 0.8. We constrained
the PDS normalization factor to vary in the range 0.7-0.9, to allow
for confusion in the PDS on-source and off-source collimators due to
faint hard sources.  The energy range used for the fits are: 0.1-4 keV
for the LECS (channels 12-400), 1.65-10 keV for the MECS (channels
37-220) and 13-200 keV for the PDS. Errors quoted in this paper are
90\% confidence intervals for two interesting parameters, unless 
differently specified.

\section{Results}

\begin{figure}
\centering
\includegraphics[angle=-90,width=8cm]{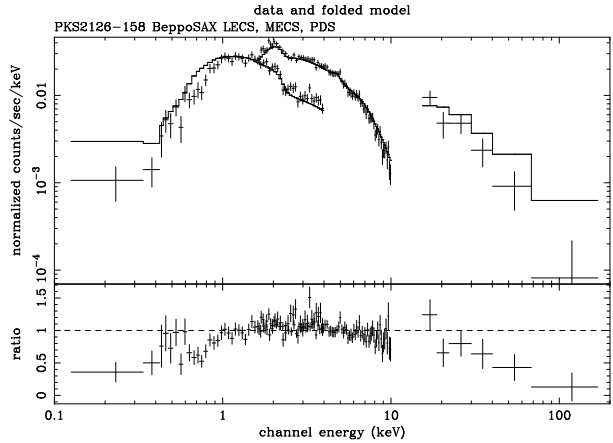}
\caption{The LECS, MECS and PDS spectra of PKS2126-158 fitted with
a power law model with Galactic absorption. }
\label{spe1}
\end{figure}

Figure \ref{spe1} shows the LECS+MECS+PDS spectra fitted with 
a power law model absorbed by the Galactic column density
along the line of sight ($4.85\times 10^{20}$ cm$^{-2}$
\footnote {Cappi et al. (1997) verified that large
molecular clouds are not present along the line of sight to
PKS2126-158, and that the Galactic gas column density estimated
from the IRAS $100\mu$ maps, assuming a typical gas to dust ratio,
is $\ls$ than the 21cm estimate.} Elvis et al. 1989). 
The figure shows that the 0.1-100 keV spectrum of PKS2126-158
is complex. A strong cut-off below 1 keV is clearly visible as well 
as a broad-band spectral curvature.

To allow an easier comparison with results obtained with other
satellites we give in table 2 the results of a fit with a simple power
law model reduced at low energy by the Galactic column density of
$4.85\times 10^{20}$ cm$^{-2}$ and by an additional column at zero
redshift, assuming solar abundances.  The $\chi^2$ is acceptable, but
relatively large residuals are present at both low (0.4-0.8 keV) and
high ($>20$ keV) energies.  Both the spectral index and the absorbing
column are consistent with previous ROSAT (Elvis et al. 1994) and ASCA
(Serlemitsos et al. 1994, Cappi et al.  1997, Reeves \& Turner 2000)
determinations.

\begin{figure}
\centering
\includegraphics[angle=-90,width=8cm]{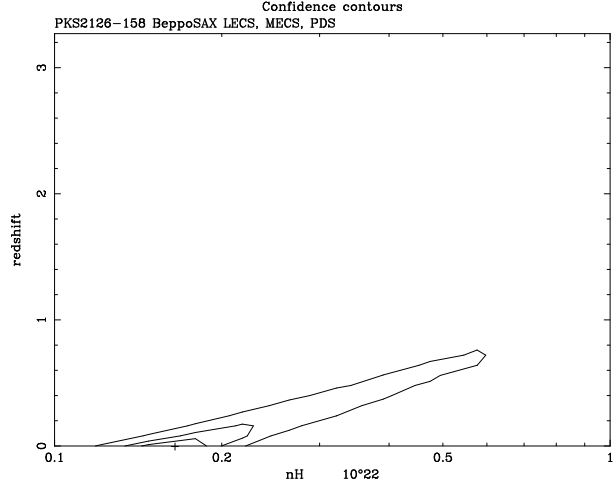}
\caption{
$N_H$ vs redshift $\chi^2$ 67\%, 90\% and 99\% 
confidence contours for a power 
law continuum.
}
\label{cont1}
\end{figure}

Allowing the redshift of the absorber to be free to vary does not
improve the fit. Figure \ref{cont1} shows the $\chi^2$ confidence 
contours for the $N_H$ and the redshift for a power law model continuum.
With this choice for the continuum shape we can conclude that absorption
in addition to the Galactic value is required by the data at high confidence
level and that the redshift of the absorber must be smaller than
0.8 at the 99\% confidence level.

\begin{figure}
\centering
\includegraphics[angle=-90,width=8cm]{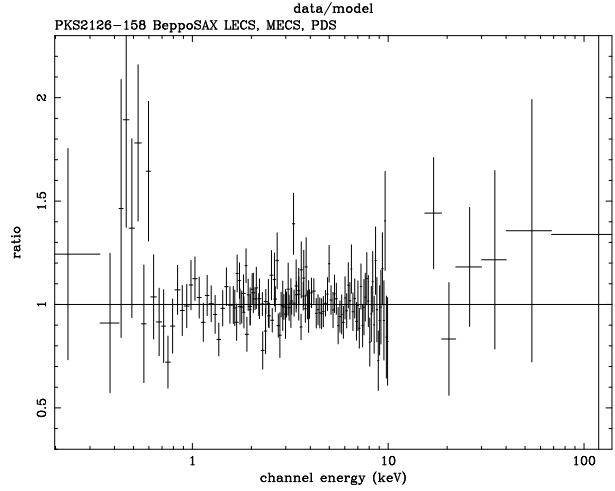}
\caption{
Ratio between the BeppoSAX LECS, MECS and PDS spectra of PKS2126-158 and
the best fitting power law with exponential cutoff model.
}
\label{ratio}
\end{figure}

\begin{figure}
\centering 
\includegraphics[angle=-90,width=8cm]{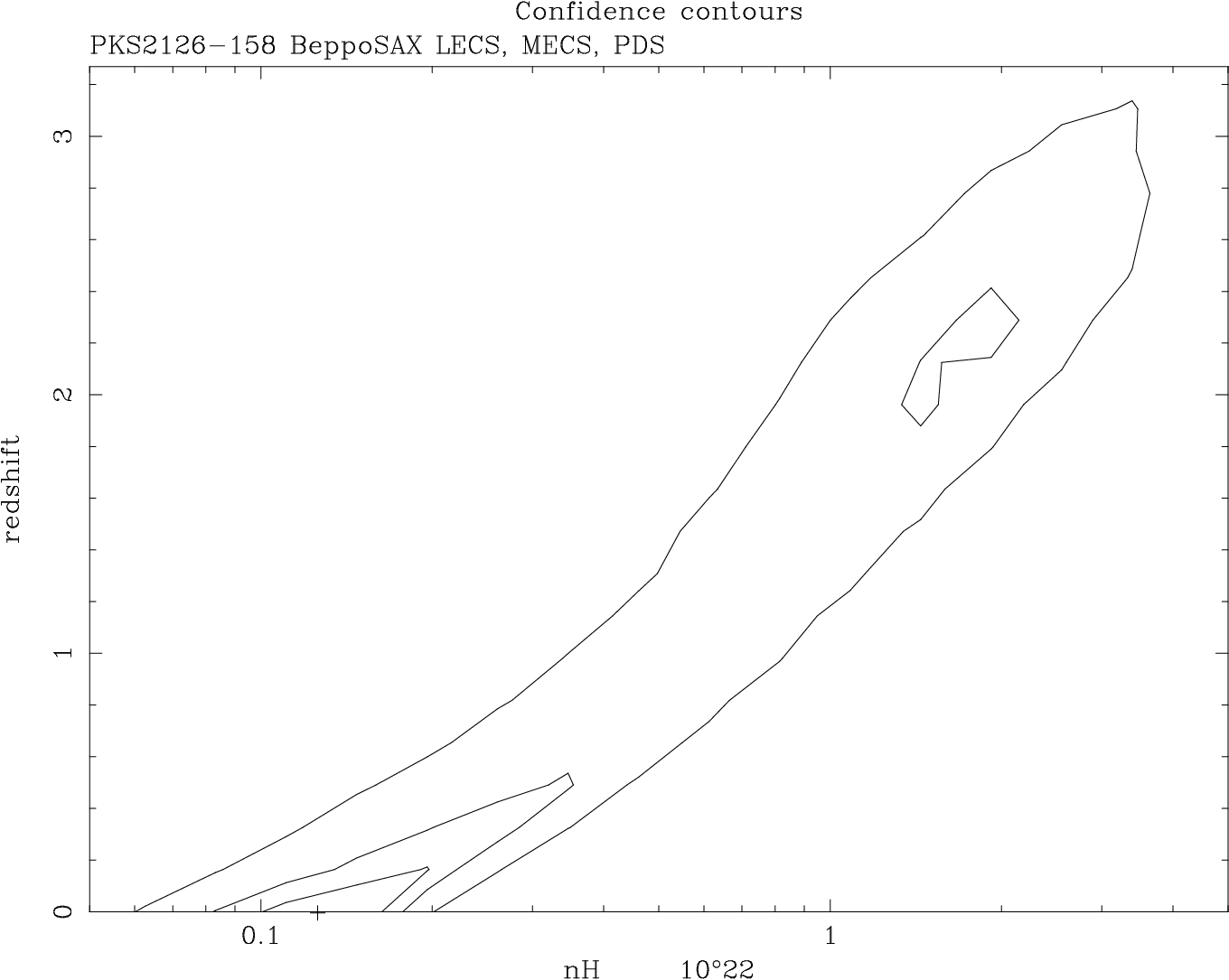}
\caption{$N_H$ vs redshift $\chi^2$ 67\%, 90\% and 99\% 
confidence contours for a broken power 
law continuum. }
\label{cont2}
\end{figure}

\begin{figure}
\centering 
\includegraphics[angle=-90,width=8cm]{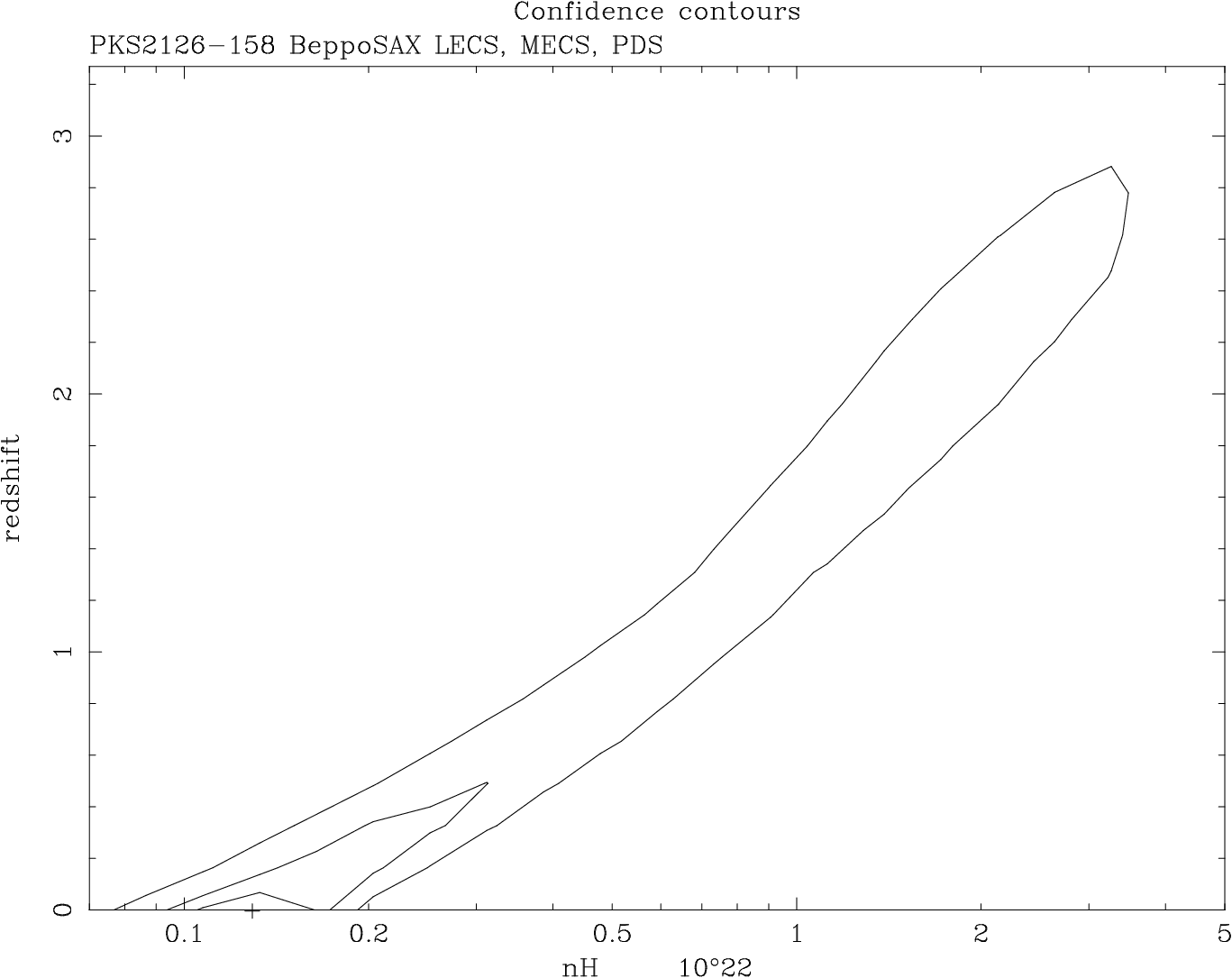}
\caption{$N_H$ vs redshift $\chi^2$ 67\%, 90\% and 99\% confidence contours 
for a power law with exponential cutoff continuum.}
\label{cont3}
\end{figure}

\begin{table*}[ht]
\caption{\bf Spectral fits with a neutral absorber}
\footnotesize
\begin{tabular}{lcccccc}

\hline
model & $N_H^a$ 	& z 	& $\alpha_E^b$ 	&$\alpha_{H}^c$	
& E$_{break}^d$ or E$_{cut}^e$	& $\chi^2$(dof) \\ 
\hline
PL	& 0.16$\pm0.03$	& 0FIX  & 0.62$\pm$0.03	&		
&		& 145.9/144\\
CUTOFFPL& 0.13$^{+0.05}_{-0.03}$ &0FIX	& 0.47$^{+0.11}_{-0.07}$& 	
& 150$^{+300}_{-50}$& 131.8/143\\
BKNPL	& 0.13$^{+0.04}_{-0.03}$ & 0FIX  & 0.73$^{+0.32}_{-0.08}$
& 1.52$\pm0.08$ & 5.0$^{+3.0}_{-1.5}$	& 136.1/142\\
Curved  & 0.13$^{+0.03}_{-0.03}$ & 0FIX  &  0.49$^{+0.09}_{-0.26}$
& 3.5$^{+3.0}_{-1.8}$  & 36$^{+19}_{-11}$  & 131.8/142\\
%BKNPL	& 1.74		&2.2FIX & 1.71		& 1.44
%& 3.6		& 139.9/142\\
%BKNPL	& 1.72		&3.27FIX& 1.60		& 1.15
%& 2.0		& 146.6/142\\
\hline
\end{tabular}

\normalsize

$^a$ in $10^{22}$ cm$^{-2}$; \\
$^b$ power law energy index, $F(E)\propto E^{-\alpha_E}$;\\
$^c$ hard power law energy index in the BKNPL and Curved models,
$F(E)\propto E^{-\alpha_E}$ for $E<E_{break}$ and  
$F(E)\propto E^{-\alpha_H} \times  E_{break}^{(\alpha_H-\alpha_E)}$
for $E>E_{break}$. The Curved model is similar to the BKNPL model
but the two power laws are joined smoothly; \\
$^d$ Break energy in keV in the BKNPL and Curved models;\\
$^e$ Exponential cut-off energy in keV in the CUTOFFPL model,
$F(E)\propto E^{-\alpha_E}\times e^{E/E_{cut}}$.

\end{table*}

However, figure \ref{spe1} indicates some curvature of the underlying
continuum. This was already suggested by ASCA observations
(Serlemitsos et al. 1994, Cappi et al. 1997).  Intrinsic curvature of
the low energy spectrum will of course result, in low resolution
spectra, in a change of the shape of the actual low energy
photoelectric cutoff and therefore of the estimated absorber
properties (density, redshift, physical and chemical composition).

We have therefore fitted the data using a power law with a high energy
exponential cutoff (see table 2). The improvement in $\chi^2$ with
respect to the simple power law model is significant at $>99.99\%$
confidence level. The ratio between the data and the best fit model is
shown in figure \ref{ratio}.  Allowing the redshift of the absorber be
free to vary does not improve the fit. Figure \ref{cont2} shows the $N_H$
vs. redshift $\chi^2$ confidence contours for the power law with
exponential cutoff model continuum.

We have also fitted the data using a broken power law model and with
with a broken power law model with a smooth conjunction between the
two power laws (``Curved'' model). The results are again in Table 2.
For the broken power law model The improvement in $\chi^2$ with
respect to the simple power law model is significant at the 99.28\%
confidence level.  Figure \ref{cont3} shows the $N_H$ vs. redshift
$\chi^2$ confidence contours for a broken power law model continuum.
There are two minima in the $\chi^2$ surface, one at z=0 and the
second at z$\sim2.2$. The best fit results for the ``curved'' model
resemble those of the power law with exponential cutoff model, but
with an additional free parameter.

Figures \ref{cont2} and \ref{cont3} show that if the absorption is
neutral an the continuum is paremeterized with either a broken power
law or a power law with a cutoff, the redshift of the absorber is less
that that of the quasar at better than the 99\% confidence level.

For the three continuum models, we have also fitted an ionized absorber
model ({\sc Absori} in {\sc XSPEC}) at both z=0 and z=3.27.  The
results are given in Table 3. The continuum parameters are very close to
those in Table 2 and therefore, for sake of simplicity, we give in
Table 3 the parameters of the absorber only.  The fits with the
absorber at z=0 give a good $\chi^2$ in all cases. Equally good fits
for an absorber at the quasar redshift can only be obtained if the
iron abundance Z$_{Fe}$
is substantially less than solar.  (The iron K edges
are redshifted into the 1.6-2.0 keV band. At these energies the quality
of the LECS and MECS spectra can strongly constrain the iron abundance.)
Since the abundance of all other elements is fixed to the solar
value, the fits actually constrain the [$\alpha$/Fe] ratio to be
higher than 0.5.

\begin{table*}[ht]
\caption{\bf Spectral fits with an ionized absorber$^a$}
\footnotesize
\begin{tabular}{lccccc}

\hline
model & $N_H^b$ 	& z 	& $\xi^c$  	& Z$_{Fe}^d$    	& 
$\chi^2$(dof) \\ 
\hline
PL	& 0.26$\pm0.05$	& 0FIX  &  0.21$^{+0.50}_{-0.20}$  & 1FIX	& 
 134.7/143\\
BKNPL	&  $0.23^{+0.20}_{-0.10}$ & 0FIX  & $<0.8$ &  1FIX 		&  
 131.4/140\\
CUTOFFPL& $0.20^{+0.10}_{-0.06}$ & 0FIX	& $<1.0$         & 1FIX		& 
 126.5/142\\
\hline
PL	& 50$^{+150}_{-25}$ & 3.27FIX 	& 1800$^{+2500}_{-1400}$ & $<0.25$&
 140.8/142\\
BKNPL	& 36$^{+150}_{-24}$ &3.27FIX& 1300$^{+3700}_{-1100}$ & $<0.25$	&
 129.5/141\\
CUTOFFPL& 36$^{+150}_{-24}$ & 3.27FIX & 1700$^{+3800}_{-1500}$ &  $<0.3$&
 127.6/141\\
\hline
\end{tabular}

\normalsize
$^a$ continuum fit values are indistinguishable from those in 
Table 2;\\
$^b$ in $10^{22}$ cm$^{-2}$; $^c$ ionization parameter;
$^d$ iron abundance in unit of the solar value.

\end{table*}

\begin{figure}[t]
\centering
\includegraphics[angle=-90,width=8cm]{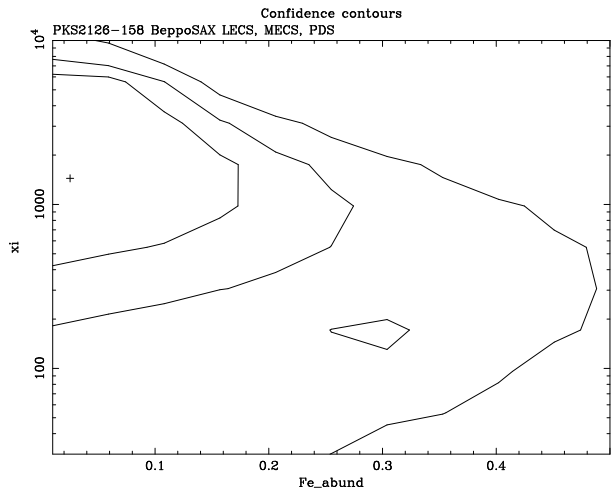}
\caption{Ionization parameter vs. iron abundance $\chi^2$ 67\%, 90\% and 99\% 
confidence contours 
for a power law with exponential cut-off continuum. 
}
\label{fexi}
\end{figure}

As an example Figure \ref{fexi} shows the $\chi^2$ confidence contours
for the iron abundance versus the ionization parameter, for a power
law continuum model, assuming that the absorber is at the redshift of
the quasar.  The analogous contours for the other two parameterizations
of the continuum are similar.  In all cases the best fit solution
implies column densities of the order of a few$\times 10^{23}$
cm$^{-2}$, high ionization parameter and low iron abundance.  The
inclusion of an additional cold absorber at the quasar redshift does
not improve the $\chi^2$.

\section{Discussion}

\subsection{Flux, Luminosity and variability}

The 2-10 keV and 0.1-2.5 keV fluxes observed by BeppoSAX are
$1.1\times10^{-11}$ and $4.4\times10^{-12}$ \cgs respectively.  The
unabsorbed 0.1-2.5 keV flux is 2.1 times higher ($1.0\times10^{-11}$
\cgs).  The statistical uncertainties on these values is of the order
of a few \%, smaller than the systematic uncertainty in the LECS and
MECS flux absolute calibration (which is of the order of 5 \%, Sacco,
private communication, also see Fiore, Guainazzi \& Grandi 1999).  
The unabsorbed fluxes correspond to a 2-10
keV and 0.1-2.5 keV luminosities of $1.6\times10^{48}$ erg s$^{-1}$
and $1.5\times10^{48}$ erg s$^{-1}$ respectively, assuming
``isotropic'' emission and using H$_0$=50 and $\Omega_0$=0.  For a
H$_0$=65, $\Omega_0$=0.3 and $\Omega_{\Lambda}=0.7$ cosmology the 2-10
keV and 0.1-2.5 keV luminosities are 2.5 times smaller
($6.8\times10^{47}$ erg s$^{-1}$ and $6.2\times10^{47}$ erg s$^{-1}$
respectively).

\begin{figure}
\centerline{ 
\includegraphics[angle=-90,width=8cm]{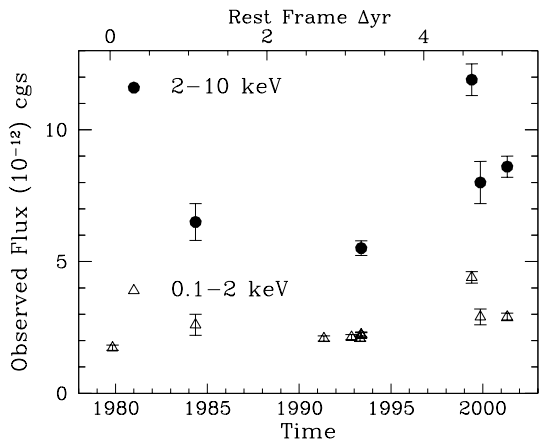}
}
\caption{2-10keV (filled circles) and 0.1-2keV (open triangles) 
fluxes observed by Einstein, EXOSAT,
ROSAT, ASCA, BeppoSAX and XMM-Newton in the last 22 years.
Error bars are 1 $\sigma$ statistical uncertainties}
\label{flux}
\end{figure}

The observed fluxes are compared in figure \ref{flux} with previous
measurements obtained with the Einstein, EXOSAT, ROSAT and ASCA
satellites (Worrall \& Wilkes 1990, Elvis et al. 1994, Cappi et
al. 1997) and with the fluxes recorded in two recent Chandra and
XMM-Newton observations (D'elia et al.  2003 in preparation, Ferrero
\& Brinkmann 2003). 5\% systematic errors have been added in
quadrature to the statistical errors on the fluxes.  Figure \ref{flux}
shows that BeppoSAX caught the source in its historical maximum, more
than a factor of 2 brighter than all previous observation and 40\%
brighter than in the later Chandra and XMM-Newton observations. A
detailed analysis of the source spectral variability among the ASCA,
BeppoSAX, Chandra and XMM-Newton observation is beyond the scope of
this paper and will be presented elsewhere (D'Elia et al. 2003).  The
rest frame time delay between the ASCA and BeppoSAX observations is
about one year, while that between the BeppoSAX and Chandra and
BeppoSAX and XMM-Newton observations is of 1.5 and 6 months
respectively.

PKS2126-158 is a GigaHertz Peaked Spectrum (GPS) quasar (O'Dea et
al. 1991, Stanghellini et al. 2001).  On a 3 miliarcsec scale the
source has core-jet morphology, with a possible counter jet detection.
Based on the possible counter jet detection and, more importantly, on
the low polarization, Stanghellini et al.  (2001) conclude that
PKS2126-158 can be only marginally beamed.  If beaming is not large in
this source, then the above ``isotropic'' luminosities are likely to
be close (within a factor of a few) to the real luminosities. These
then would imply extreme black hole masses of at least a
few$\times10^{10}$M$_{\odot}$, assuming that the quasar is emitting
close to its Eddington luminosity.  The causal timescale associated to
an emission region of $\sim10$ Schwarzschild radii around these black
holes is $\sim$months, comparable to the shortest variability
timescale detected.

The absorbing column density depends of course by the redshift.  If
z=0 is assumed then the BeppoSAX values in Table 2 are well consistent
with the ROSAT and ASCA ones. If z=3.27 is assumed then the column
densites in figures \ref{cont2} and \ref{cont3} ($2-4\times10^{22}$
cm$^{-2}$), are somewhat higher than the ROSAT and ASCA values (Elvis
et al. 1994, Cappi et al. 1997), but still statistically compatible
with them. We note that this little discrepacy is likely due to the
fact that the continuum assumed by the latter authors is a simple
power law, while figures \ref{cont2} and \ref{cont3} are obtained
using a broken power law and a power law with exponential cutoff
continuum.

\subsection{Low Energy absorption}

The BeppoSAX NFI observations of PKS2126-158 confirm the presence of
low energy absorption in this source {\it independent} of the
continuum form. Unlike all previous investigations, even assuming a
curved continuum and a broken power law continuum, the spectra do
require, at a high confidence level, absorption in excess to the
Galactic column density along the line of sight.  Because of the
limited spectral resolution of the LECS and MECS it is difficult to
distinguish between different absorption and continuum models and to
assess precisely the redshift of the absorber.  If a power law model
is used to parameterize continuum the redshift of the absorber is
smaller than 0.8 at the 99\% confidence level.  If the continuum is
parameterized using a broken power law the redshift of the absorber is
more poorly constrained (the 99\% $\chi^2$ contours nearly touch the
emission redshift of 3.27). Interestingly there are two minima in the
$\chi^2$ surface, one at z=0 and the second at z=2.2.  If the
continuum is parameterized using a power law with an exponential
cutoff then z$_{abs}<3$ at the 99\% confidence level.  In summary, if
the absorber is not strongly ionized, the BeppoSAX data do prefer an
absorber at z$<$z$_{em}$.  X-ray absorption up to a level of $10^{22}$
cm$^{-2}$ can be produced by DLA systems along the line of sight. This
is seen in several low-z DLAs by Chandra (Bechtold et al. 2001,
Siemiginowska et al. 2002).  Although, no DLA is observed toward
PKS2126-158, only a small part of the redshift range is accessible to
direct DLA searches, since a z=2.9676 Lyman limit system cuts out
shorter wavelength UV emission, so DLAs cannot be directly detected
below  z=1.976.  However, several complex metal line systems have
been detected at 0.66$<$z$<$2.9 (Giallongo et al. 1993, D'odorico et
al. 1998).  In particular, the system at z=0.6631 has strong low
ionization CaII3934-3969 and MgII2796-2803 absorption. The MgII column
is $4(\pm1)\times10^{17}$ cm$^{-2}$ (D'odorico et al. 1998), which
would correspond to a hydrogen column of $\sim10^{22}$ cm$^{-2}$
for solar abundances, assuming that the MgII ion is the most populated
of the Mg ions. Since other Mg ions will likely to be populated the
hydrogen column could actually be higher than this value.  Therefore
a DLA could well be associated to the 0.6631 system and could be
responsible for the X-ray absorption. Strong, $\tau>1$, low
ionization oxygen edges at $\sim0.33$ keV are predicted if
$N_H\gs10^{22}$ cm$^{-2}$ and if the gas ionization is as low, as
suggested by the strong MgII line.  Furthermore, at z=2.63-2.82 there
are 5 metal systems within a velocity range of 800 km/s
(Giallongo et al. 1993).  They clearly indicate a rich environment at
this redshift in the direction of PKS2126-158, which could again well
be associated with the site of the X-ray absorption.

%Although the source was a factor of two brighter during the BeppoSAX
%observations than in all previous ASCA and ROSAT observations, the
%level of absorption remained roughly the same, as if the absorber has
%not ``responded'' to the continuum variation on the same
%timescale. This could be the case if the variability timescale is
%smaller than the gas photoionization timescale, i.e. the time that the
%gas takes to reach a new photoionization equilibrium.  When the ion
%abundances are distributed mainly between two ionic species the
%photoionization timescale of these particular ionic species is
%inversely proportional to the gas density times a function of the
%ratios between the abundances in the two ionic species (Nicastro et
%al. 1999a,b). In the warm absorber solution for PKS2126-158 oxygen is
%distributed mainly between OVIII and OIX.  The OVIII absorption edge and
%resonant line will be redshifted below $\sim0.2$ keV, at energies where 
%unfortunately the
%Galactic column density strongly reduces the observed spectrum.
%From the upper limit to the
%variability timescale we can therefore obtain an upper limit to the
%gas density of a few$\times 10^3$ cm$^{-3}$.  From the definition of
%ionization parameter this translates to a limit on the distance of the
%absorbing gas from the ionizing source of greater than 10 pc.  

On the other hand, a highly ionized absorber at the quasar redshift
provides a good fit only if the iron abundance is smaller than $\sim0.3$
solar, while those of all other elements are fixed to the solar value.
The ionized absorber could be the result of the interaction of the radio
jet of PKS2126-158 with the ISM of the host galaxy.  PKS2126-158 is a
GPS quasar and GPS
sources are often explained in term of a young, newly born radio
sources.  Silk \& Rees (1998) and Fabian (1999) proposed a scenario in
which newly born, high luminosity quasars, are still enshrouded by
large densities of soft X-ray absorbing gas. If its X-ray absorption
is intrinsic PKS2126-158, could then be an example of these newly
born, highly obscured quasars.  The connection of X-ray cut-offs and
newly born quasars is highly suggestive, and it must be studied in
detail with future high resolution X-ray observation, which can be
able to constrain tightly the redshift of the absorbers.

In the ionized absorber solution the limit to the iron abundance is
rather tight (Z$_{Fe}<0.3$).  This is due to the good sensitivity of
the LECS and MECS at 1.5-2 keV, where the Fe-K edges are
redshifted. This limit contrasts with the estimate of the nuclear iron
abundance obtained by Thompson et al. (1999) using the intensity of
iron optical emission lines.  However, as mentioned in the
introduction optical estimates may be inaccurate.  The X-ray
metallicity determination and optical lines measurements could be
combined to constrain the line emitting gas physical status.  An
underabundance of iron ([$\alpha$/Fe]$>0.5$), would be consistent
with a scenario in which the ISM has been enriched mostly by type II
SNe. If confirmed, this could be used to put constraints on the time
needed to form the bulk of type Ia SNe, i.e the time to form a WD in a
binary system plus the time needed to accrete matter from the
companion to surpass the Chandrasekar limit, which brings to the SN
explosion. All these time-scales are today rather poorly defined.

Since iron aggregates easily in dust grains, an iron abundance of the
X-ray absorbing matter significantly less than the solar value can
contribute to explaining the lack of strong ultraviolet reddening in
this source (Elvis et al. 1994, Bechtold et al. 1994), expected from
the measured X-ray absorbing column density, if the dust to gas ratio
and dust composition are similar to the Galactic one.  A low iron
abundance absorber can also help in solving the apparent paradox of
hard X-ray selected high redshift quasars which appear to be obscured
in X-rays but show strong broad optical and UV lines and little
extinction in the optical band (see Fiore et al. 2001 and Comastri et
al. 2001, Maiolino et al. 2001a).  Other possible solutions of these
contradictions are: (1) the X-ray absorber could be highly ionized and
the dust destroyed, if the matter is close enough to the active
nucleus; (2) the dust grains could have different properties with
respect to the ``standard'' Galactic mixture (Maiolino et al. 2001b).

Although the ionized absorber hypothesis is statistically tenable and
astrophysically intriguing, in that it helps in resolving the
contradiction between the X-ray absorption and the lack of UV
extinction, the present low resolution data are inadequate to choose
this as a unique solution.  The precise determination of the absorber
redshift, as well as of its metallicity and ionization state must
await for the high resolution observations that have become possible
with the gratings on board of the Chandra and XMM-Newton satellites.

%%%%%%%

\begin{acknowledgements}
We would like to thank the BeppoSAX hardware teams
for the development and calibration of the instruments.  In
particular, we would like to remember the late Daniele Dal Fiume for
his continuous dedication and his fundamental contribution to an
instrument of unprecedented sensitivity like the
BeppoSAX PDS.  We thank Emanuele Giallongo for useful discussions.
This research has made use of SAXDAS linearized and cleaned event
files (Rev.2.0) produced at the BeppoSAX Science Data Center.  This
research has been partially supported by ASI contract ARS--99--75 and
MURST grant Cofin--98--032.

This research has made use of the NASA/IPAC Extragalactic
Database (NED) which is operated by the Jet Propulsion
Laboratory, Caltech, under contract with the National Aeronautics
and Space Administration, and the HEASARC Databases.
\end{acknowledgements}
%%%%%%%

%%%%%%%%%%%%%%

\begin{thebibliography}{}

\bibitem[]{}Arav et al. 2001, ApJ, 561, 118

\bibitem[]{}Bechtold, J. et al. 1994, AJ, 108, 374

\bibitem[]{}Bechtold, J., Siemiginowska, A., Aldcroft, T.L., Elvis, M.,
\& Dobrzycki, A. 2001, ApJ, 562, 133

\bibitem[]{}Boella, et al., 1997a, A\&AS, 122, 299

\bibitem[]{}Boella, et al., 1997b, A\&AS, 122, 327

\bibitem[]{}Cappi, M. et al. 1997, ApJ, 478, 492

\bibitem[]{}Churchill, C.W., Mellon, R.R., Charlton, J.C., Jannuzi, B.T.,
Kirhakos, S., Steidel, C.C., Schneider, D.P. 2000, ApJS, 130, 91

\bibitem[]{}Constantin, A., Shields, J.C., Hamann, F., Folz, C.B., 
Chaffee F.H. 2002, ApJ, 565, 50

\bibitem[]{}Dietrich, M., Appenzeller, I., Vestergaard, M., Wagner, S.J. 2002,
ApJ, 564, 581

\bibitem[]{}D'Odorico, V., Cristiani, S., D'Odorico, S., Fontana, A.
Giallongo, E. 1998, A\&AS, 127, 217

\bibitem[]{}Elvis, M., Lockman, F.J. \&  Wilkes, B.J. 1989, AJ, 97, 777

\bibitem[]{}Elvis, M., Fiore, F., Wilkes, B.J., McDowell, J.C. \& Bechtold, J.,
1994a, ApJ, 422, 60.

\bibitem[]{}Elvis, M., Fiore, F., Giommi, P. \& Padovani, P., 1998, ApJ, 492, 91.

\bibitem[]{}Fabian, A.C. 1999, MNRAS, 308, L39

\bibitem[]{}Ferrero, E. \& Brinkmann, W. 2003, A\&A in press, astro-ph/0302421

\bibitem[]{}Fiore, F., Elvis, M., Giommi, P. \& Padovani, P., 1998, ApJ, 492, 79.

\bibitem[]{}Fiore, F., Guainazzi, M. \& Grandi, P. 1999,
Handbook for BeppoSAX NFI spectral analysis,
ftp://ftp.asdc.asi.it/pub/sax/doc/software\_docs/saxabc\_v1.2.ps.gz or
http://heasarc.gsfc.nasa.gov/docs/sax/abc/saxabc/ 

\bibitem[]{}Frontera F. et al., 1997, A\&AS, 112, 357

\bibitem[]{}Giallongo, E., Cristiani, S., Fontana, A., Trevese, D. 1993, 
ApJ, 416, 137

\bibitem[]{}Hamann, F. 1998, ApJ, 500, 798

\bibitem[]{}Hamann, F. \& Ferland, G.J. 1999, ARAA, 37, 487

\bibitem[]{}Hamann, F., Korista, K.T., Ferland, G.J., Warner, C., 
Baldwin, J. 2002, ApJ, 564, 592

\bibitem[]{}Hamann, F., et al. 2003, proc. of the workshop ``X-ray
spectroscopy of AGN with Chandra and XMM'', Garching, Germany December 2001

\bibitem[]{}Kuhn, O., Elvis, M., Bechtold, J., Elston, R. 2001, ApJS, 
136, 225

\bibitem[]{}Iwamuro, F., Motohara, K., Maihara, T., Kimura, M.,
Yoshii, Y., Doi, M. 2002, ApJ, 565, 63

\bibitem[]{}Levine, A.M. et al., 1984, ApJS, 54, 581.

\bibitem[]{}Madejski, G., Takahashi, T., Tashiro, M., Kubo, H.,
Hartman, R., Kallman, T., \& Sikora, M. 1996, ApJ., 459, 156

\bibitem[]{}Maiolino, R., Marconi, A., Salvati, M.,
Risaliti, G., Severgnini, P., Oliva, E.,
La Franca, F., Vanzi, L. 2001a, A\&A, 365, 28

\bibitem[]{}Maiolino, R., Marconi, A., Oliva, E. 2001b A\&A, 365, 37

\bibitem[]{}Nicastro, F., Fiore, F., Perola, G.C, Elvis, M. 1999a, ApJ, 512, 184

\bibitem[]{}Nicastro, F., Fiore, F., Perola, G.C, Elvis, M. 1999b, ApJ, 512, 136

\bibitem[]{}O'Dea, C.P., Baum, S.A., Stanghellini, C. 1991 ApJ, 380, 66 

\bibitem[]{}Parmar A. et al., 1997, A\&AS, 112, 309

\bibitem[]{}Pettini, M., Smith, L.J., King, D.L., Hunstead, R.W. 1997, 
ApJ, 486, 665

\bibitem[]{}Pettini, M., Ellison, S. L., Steidel, C.C., Bowen, D.V.  
1999, ApJ, 510, 576

\bibitem[]{}Prochaska J. X., Wolfe, A.M. 1999, ApJS, 121, 369

\bibitem[]{}Reeves, J.N., \& Turner, M.J.L. 2000, MNRAS, 316, 234

\bibitem[]{}Savaglio S., Panagia, N.,  Stiavelli, M. 2000,  
ASP Conference Proceedings, 
Vol. 215, ed. J. Franco, L. Terlevich, O. Lopez-Cruz, and I. Aretxaga. 
Astronomical Society of the Pacific,  p.65. Astro-ph/0011473

\bibitem[]{}Serlemitsos, P. et al. 1994, PASJ, 46, L43

\bibitem[]{}Silk, J. \& Rees, M. 1998, A\&A, 331, L1

\bibitem[]{}Stanghellini, C., Dallacasa, D., O'Dea, C.P., Baum, S.A., 
Fanti, R., \& Fanti, C., 2001, A\&A, 377, 377

\bibitem[]{}Thompson, K.L., Hill, G.J., \& Elston, R. 1999, 515, 487

\bibitem[]{}Verner E. M., Verner, D. A., Korista, K. T.,
Ferguson, J. W., Hamann, F., Ferland, G. J. 1999, ApJS, 120, 101

\bibitem[]{}Worral, D.M., \& Wilkes, B.J. 1990, ApJ, 360,396 

\end{thebibliography}
\end{document}